\documentclass[preprintnumbers,showpacs,amsmath,amssymb,floatfix,prd,onecolumn,superscriptaddress,nofootinbib]{revtex4}
\usepackage{amssymb}
\usepackage{bbm}
\usepackage{graphicx}
\usepackage{epsfig}
\usepackage{bm}
\usepackage{amsfonts}
\usepackage{epstopdf}
\usepackage{multirow}

\begin{document}

\title{Casimir Effect under  Quasi-Periodic Boundary Condition Inspired by Nanotubes}

\author{Chao-Jun Feng \footnote{Corresponding author.}}
\email{fengcj@shnu.edu.cn} \affiliation{Shanghai United Center for Astrophysics (SUCA), \\ Shanghai Normal University,
   100 Guilin Road, Shanghai 200234, P.R.China} 
   \affiliation{State Key Laboratory of Theoretical Physics, \\Institute of Theoretical Physics, Chinese Academy of Sciences, Beijing 100190, P.R.China}

\author{Xin-Zhou Li}
\email{kychz@shnu.edu.cn} \affiliation{Shanghai United Center for Astrophysics (SUCA),  \\ Shanghai Normal University,
    100 Guilin Road, Shanghai 200234, P.R.China}

\author{Xiang-Hua Zhai}
\email{zhaixh@shnu.edu.cn} \affiliation{Shanghai United Center for Astrophysics (SUCA),  \\ Shanghai Normal University,
    100 Guilin Road, Shanghai 200234, P.R.China}

\begin{abstract}
When one studies the Casimir effect, the periodic (anti-periodic) boundary condition is usually taken  to mimic a periodic (anti-periodic) structure for a scalar field living in a flat space with a non-Euclidean topology.  However, there could be  an arbitrary  phase difference between the value of the scalar field on one endpoint of the unit structure and that on the other endpoint, such as the structure of nanotubes.  Then, in this paper, a periodic condition on the ends of the system with an additional phase factor, which is called the ``quasi-periodic" condition , is imposed to investigate the corresponding Casimir effect. And an attractive or repulsive Casimir force  is found, whose properties  depend on the phase angle value. Especially, the Casimir effect disappears when the phase angle takes a particular value.  High dimensional space-time case is also investigated.
\end{abstract}

\maketitle

\flushbottom


\section{Introduction}\label{sec:intro}

The dynamics of a classical or a quantum field drastically depends on the external boundary conditions imposed on it.  The spectrum of the field could be changed with a boundary condition, and it will lead to an observable effect. Casimir effect \cite{Casimir}\cite{Plunien:1986ca} is one of such kind of phenomena. In its simplest form, the presence of two infinitely large, perfectly reflecting parallel conducting planes placed in a vacuum changes the spectral density of zero-point fluctuation of the vacuum and leads to an attractive force between the planes.  This is an entirely quantum effect because the force acting between two neutral plances is equal to zero in classical electrodynamics.

Since the last decade, the Casimir effect has been paid more attention due to the development of precise measurements \cite{Decca:2007yb} and technological advancements. Indeed, the Casimir force offers new possibilities for nanotechnology, such as the  actuation of micro- and nanoelectromechanical systems (MEMS and NEMS) mediated by the quantum vacuum \cite{MEMS, MEMS2}. However, there are some challenges since the same force could be generated by the stiction of these devices and there also a lot of work both from theoretical and experimental aspects to understand how to engineer the strength and sign of the Casimir force - a repulsive force would provide an anti-``stiction'' effect \cite{phan,phan2}.   Inevitably a lot of attention has been focused on the role of boundary conditions and very  recently on the interplay of material properties, temperature, and geometry.  Some new methods have developed for computing the Casimir effect between a finite number of compact objects \cite{Emig:2007cf},  inside a rectangular box or cavity \cite{Li, Li2, Li3}.  When  a topology of the flat spacetime was chosen to  cause the helix boundary condition for a  scalar field,  the Casimir force behaves very much like the force on a spring that obeys the Hooke's law when the ratio of the pitch to the circumference of the helix is small, but in this case, the force comes from a quantum effect, so the author call it \textit{quantum spring} \cite{Feng:2010qj} \cite{Zhai:2010mr} or \textit{quantum anti-spring} \cite{Zhai:2011zza} corresponding to periodic-like and anti-periodic-like boundary condition, see also \cite{Zhai:2011pt}. For recent review on the Casimir effect, see \cite{Brevik:2012ht}.

In this paper, we will present our results for the computation of Casimir energies and forces  for a scalar mimicking  a periodic nanostructure by means of  zeta function regularization method \cite{Elizalde}.  The nanostructure under considered could be a $2D$ lamellar grating, which is a periodic metallic and/or dielectric structure that consist of planar layers.  The $\zeta$-function
regularization procedure is a very powerful and elegant technique for the Casimir effect. Rigorous extension of the
proof of Epstein $\zeta$-function regularization has been discussed in \cite{Elizalde}. Vacuum polarization in the
background of on a string was first considered in \cite{Helliwell:1986hs}. The generalized $\zeta$-function has many
interesting applications, e.g., in the piecewise string \cite{Li:1990bz, Li:1990bz2}. Similar analysis has been applied to noncommutative spacetime \cite{Teo:2010hr}, monopoles \cite{BezerradeMello:1999ge}, p-branes \cite{Shi:1991qc} or pistons \cite{Zhai, Zhai2, Zhai3, Zhai4, Zhai5}. Casimir effect for a fractional boundary condition is of interest in considering, for example, the finite temperature Casimir effect for a scalar field with fractional Neumann conditions \cite{Eab:2007zz}, while the repulsive force from fractional boundary conditions has been studied \cite{Lim:2009nk}.

Usually, one will take the (anti-) periodic boundary condition for the scalar field  to mimic the external boundary conditions (such as the lamellar grating ) or some non-Euclidean  topologies (like $S^{1}$). 
\begin{equation}
	\phi(t, \mathbf{x}+ \mathbf{a}) = \pm \phi(t, \mathbf{x}) \,.
\end{equation}
 However, there could be an arbitrary  phase difference between $\phi(t, \mathbf{x}+ \mathbf{a}) $ and $ \phi(t, \mathbf{x})$, namely, 
 \begin{equation} \label{bpcond}
	\phi(t, \mathbf{x}+ \mathbf{a}) = e^{i2\pi \theta} \phi(t, \mathbf{x})  \,,
\end{equation}
where the phase angle $0\leq\theta\leq 1$ and it will reduce to the (anti-) periodic boundary condition when $\theta$ takes an (half-) integer value. Generally, the phase could be any values besides $-1$ and $1$ in the complex plane. For instance, when one considers the Casimir effect in nanotubes or nanoloopes for a quantum field, $\theta = 0$ corresponds to metallic nanotubes, while $\theta = \pm 2\pi/3$ corresponds to semiconductor nanotubes. So, it is more reasonable and interesting to  take the this kind of ``quasi-periodic" boundary condition (\ref{bpcond}) for the scalar field and we found that the Casimir force could be attractive, repulsive or vanished depends on the values of $\theta$.
 
This paper is organized as follows. In the next section, the calculation of the Casimir energy and force under the quasi-periodic boundary condition for a massless and massive scalar field in $D+1$ dimensional spacetime will be presented. In the last section, we will discuss  results and prospects for future studies.

\section{Evaluation of the Casimir energy and force for a scalar field}\label{sec:casimir}
First, we consider a massless scalar field living in a flat space-time, whose dynamics is determined by the following Klein-Gordon equation
\begin{equation}\label{KG}
    \left(\partial_t^2 - \partial_i^2\right)\phi(t, x, x^{T}) = 0 \,,
\end{equation}
which has a solution as the following
\begin{equation}\label{modes}
    \phi_{n}(t, x)= \mathcal{N} e^{-i\omega_nt+ik_{x} x + ik_{T}x^{T} }\,,
\end{equation}
where $\mathcal{N}$ is a normalization factor. By taking the  quasi-periodic boundary condition (\ref{bpcond}), we get its energy spectrum
\begin{equation}\label{energy}
    w_n^2 = k_{T}^2 + \left[\frac{2\pi (n+\theta)}{a}  \right]^{2}  \,.
\end{equation}
In the ground state (vacuum state), each of these modes contributes an energy of $w_n/2$. Then the energy density of the field  in $D+1$ dimensional spacetime is given by
\begin{equation}
	E^{D}(a) = \frac{1}{2 a}\int_{-\infty}^{\infty} \frac{d^{D-1}k}{(2\pi)^{D-1}} \sum_{n=-\infty}^{\infty}w_{n}\,.
\end{equation}
In order to use the $\zeta$-function regularization, we define the function $\mathcal{E}(s)$ as
\begin{equation}\label{es}
  \mathcal{E}(a;s) =  \frac{\pi}{a^{D+1}} \sum_{n=-\infty}^{\infty}   \left(n+\theta \right)^{D-1-s}
   \int_{-\infty}^{\infty} d^{D-1}k\left(k^2 + 1\right)^{-s/2} \,,
\end{equation}
for $Re(s)>1$ to make a finite result provided by the $k$ integration. In the following one can see that its analytic continuation to the complex $s$ plane is well defined at $s=-1$. So, the regularized vacuum energy  could be written as $E^{D}_{R} = \mathcal{E}(a; -1)$.

\begin{figure}[h]
\begin{center}
\includegraphics[width=0.5\textwidth]{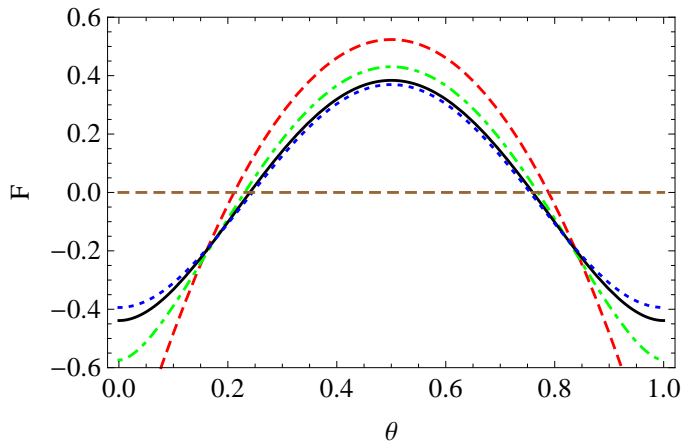}
\caption{\label{fig::forcend}The Casimir force (in unit of $a=1$) with respect to $\theta$ (in unit of $2\pi\times$rad. ) in $D+1$ dimension. Here we have plotted $D=1, 2, 3, 4$ corresponding to the dashed, dotdashed, solid and dotted curves respectively in the figure.}
\end{center}
\end{figure}

By using the mathematical identity
\begin{equation}\label{ide1}
	\int_{-\infty}^{\infty} f(x) d^{d} x = \frac{2\pi^{\frac{d}{2}}}{\Gamma(\frac{d}{2})} \int_{0}^{\infty} r^{d-1}f(r)dr \,,
\end{equation}
and the relation
\begin{equation}\label{ide2}
	\int_{0}^{\infty} t^{r}(1+t)^{s} dt = B(1+r, -s-r-1) \,,
\end{equation}
we get
\begin{equation}\label{energys}
	\mathcal{E}(a;s) = \frac{\pi^{\frac{D+1}{2}}}{a^{D+1}} \frac{\Gamma(\frac{s+1-D}{2})}{\Gamma(\frac{s}{2})}\sum_{n=-\infty}^{\infty}  \left(n+\theta \right)^{D-1-s} \,.
\end{equation}
Then using the following reflection relation, see App.~\ref{reg1},
\begin{equation}
  \pi^{-s/2}\Gamma\left(\frac{s}{2}\right)\sum_{n=-\infty}^{\infty}(n+\nu)^{-s}
	= \pi ^{-\frac{1}{2}+\frac{s}{2}} \Gamma\left(\frac{1-s}{2}\right)2\sum_{n=1}^{\infty} n^{s-1} \cos(2\pi  n \nu) \,,
\end{equation}
we  obtain the final results
\begin{eqnarray}
	\mathcal{E}(a;s) = \frac{2\pi^{s+1-\frac{D}{2}}}{a^{D+1}}  \frac{\Gamma(\frac{D-s}{2})}{\Gamma(\frac{s}{2})}\sum_{n=1}^{\infty} \frac{\cos(2\pi  n \theta)}{n^{D-s}} \,,
\end{eqnarray}
and
\begin{equation}\label{vEnergy}
	E^{D}_{R}(a) =  -\frac{\Gamma(\frac{D+1}{2})}{\pi^{\frac{D+1}{2}}a^{D+1}}\sum_{n=1}^{\infty} \frac{\cos(2\pi  n \theta)}{n^{D+1}} \,.
\end{equation}
In particular, for odd values of $D=2j+1$, ($j=0, 1, 2, \cdots $), one can get
\begin{equation}\label{vEnergyodd}
	E^{2j+1}_{R}(a) =  \frac{\Gamma\left(-j-\frac{1}{2}\right) \pi^{j+\frac{1}{2} }  \varphi_{2j+2}(\theta) }    { 2(j+1)a^{2(j+1)}} \,,
\end{equation}
where $\varphi_{n}(x)$ is the  Bernoulli polynomials, see App.~\ref{ovd}. As an example, we take $D=3$, then the energy density is given by
\begin{equation}\label{vEnergy3}
	E^{3}_{R}(a) =\frac{ \pi^{2 } }    { 3a^{4}} \bigg(-\frac{1}{30} +\theta^{2}  -2\theta^{3} +\theta^{4}  \bigg) \,,
\end{equation}
where we have used $\varphi_{4}(x) = 1/30 + x^{2}-2x^{3} + x^{4}$. The Casimir force could be derived as $F^{D}(a) = - \partial E^{D}_{R}(a)/\partial a$. For $D=3$, it becomes
\begin{equation}
	F^{3}(a) = \frac{ 4\pi^{2 } }    { 3a^{5}} \bigg(-\frac{1}{30} +\theta^{2}  -2\theta^{3} +\theta^{4}  \bigg) \,,
\end{equation}
 and we also plot the behavior of the Casimir forces in Fig.~\ref{fig::forcend}.

From Fig.~\ref{fig::forcend}, one can see that the Casimir force could be attractive or repulsive depending on values of the phase angle $\theta$ and especially, when
\begin{equation}
\theta = \frac{1}{2} \pm \frac{1}{2}\sqrt{1-\frac{2\sqrt{30}}{15}}\,,
\end{equation}
the Casimir effect disappears. The maximum value of the repulsive force is obtained when $\theta = 1/2$, which corresponds to  the anti-periodic boundary condition.  Fig.~\ref{fig::forcend} illustrates the Casimir forces in $D+1$ dimensional spacetime with  $D=1,2,3,4$ and one can see that all of them have a maximum value when $\theta = 1/2$. Actually, from eq.~(\ref{vEnergy})  we have $\partial_{\theta} F \sim \sum \sin(2\pi n \theta)/n^{D}$, then $\theta = 1/2$ makes $\partial_{\theta} F|_{\theta=1/2} = 0$ and $\partial_{\theta}^{2} F |_{\theta=1/2} < 0$, so it gets maximum value  $F_{max} = F|_{\theta=1/2}$.

In the case of  a massive scalar field, we have
\begin{equation}\label{energy}
    w_n^2 = k_{T}^2 + \left[\frac{2\pi (n+\theta)}{a}  \right]^{2} + m^{2}  \,.
\end{equation}
and
\begin{equation}
	E^{D}(a, m) = \frac{1}{2 a}\int_{-\infty}^{\infty} \frac{d^{D-1}k}{(2\pi)^{D-1}} \sum_{n=-\infty}^{\infty}w_{n}\,.
\end{equation}
To use the $\zeta$-function regularization, we define
$\mathcal{E}(s)$ as
\begin{equation}\label{es2}
  \mathcal{E}(a, m ;s) =  \frac{\pi}{a^{D+1}} \sum_{n=-\infty}^{\infty}   \bigg[ (n+\theta)^{2} + \tilde m^{2}\bigg]^{\frac{D-1-s}{2}}
	 \int_{-\infty}^{\infty} d^{D-1}k\left(k^2 + 1\right)^{-s/2} \,,
\end{equation}
for $Re(s)>1$ still to make a finite result provided by the $k$ integration and $E^{D}_{R} = \mathcal{E}(a; -1)$. Here, we have defined $\tilde m = ma/(2\pi)$ and then we have
\begin{equation}\label{energys}
	\mathcal{E}(a;s) = \frac{\pi^{\frac{D+1}{2}}}{a^{D+1}} \frac{\Gamma(\frac{s+1-D}{2})}{\Gamma(\frac{s}{2})}\sum_{n=-\infty}^{\infty}   \bigg[ (n+\theta)^{2} + \tilde m^{2}\bigg]^{\frac{D-1-s}{2}}\,.
\end{equation}
By using the following reflection relation, see App.~\ref{reg2},
\begin{eqnarray}
\nonumber
	&& \pi^{-s/2}\Gamma\left(\frac{s}{2}\right) \,  \sum_{n=-\infty}^{\infty}  \bigg[ (n+\nu)^{2} + \mu^{2}\bigg]^{\frac{-s}{2}}  \\
	&=& \pi^{\frac{1-s}{2}}\mu^{1-s}\Gamma\left(\frac{s-1}{2}\right)
+ 4 \sum_{n=1}^{\infty}\left(\frac{\mu}{n}\right)^{\frac{1-s}{2}}\cos(2\pi  n \nu)  K_{\frac{1-s}{2}}(2\pi n\mu) \,,
\end{eqnarray}
we get
\begin{eqnarray}
	&& \mathcal{E}(a, m;s) = \frac{\pi^{\frac{s}{2} + 1}}{a^{D+1}\Gamma(\frac{s}{2})} \bigg[  \pi ^{\frac{D-s}{2}}\tilde m^{D-s}\Gamma\left(\frac{s-D}{2}\right) \\
	&& \qquad +4\sum_{n=1}^{\infty}\left(\frac{\tilde m}{n}\right)^{\frac{D-s}{2}}  \cos(2\pi  n \theta) K_{\frac{D-s}{2}}(2\pi n\tilde m)
	\bigg]\,,
\end{eqnarray}
and
\begin{equation}
  E^{D}_{R}(a, m) =   -\frac{m^{D+1} \Gamma\left(-\frac{D+1}{2}\right)}{2^{D+2}\pi^{\frac{D+1}{2}}} - \frac{2}{a^{D+1}} \sum_{n=1}^{\infty}\left(\frac{\tilde m}{n}\right)^{\frac{D+1}{2}} \cos(2\pi  n \theta) K_{\frac{D+1}{2}}(2\pi n\tilde m) \,.
\end{equation}
The first contribution on the right-hand side of the above equation is associated with a constant energy density throughout the volume, and may be cancelled by a constant term in the Hamiltonian density, and it does not contribute to the corresponding Casimir force \cite{Ambjorn:1981xw}. Thus, we get the force as $F^{D}(a, m) = - \partial E^{D}_{R}(a, m)/\partial a$, which will reduce to the expression of the  force in the massless case when $m\rightarrow 0$. If the mass is large, we have
\begin{equation}
	F^{D}(a, m) \approx - \frac{(D+1) \tilde m^{\frac{D}{2}}}{a^{D+2}}
	\sum_{n=1}^{\infty}\frac{\cos(2\pi  n \theta)e^{-2\pi n\tilde m}}{n^{\frac{D}{2}+1}} \,,
\end{equation}
 which  is exponentially small. It means that the Casimir force will be vanished when the mass tends to infinity. Furthermore, the Casimir force for a massive scalar field also gets its maximum value at $\theta = 1/2$ for a given $a$ and the reason is the same as that in the massless case.

 Furthermore, if the background topology is $R^{D-q}\times (S^{1})^{q}$ in a ($D+1$) flat spacetime, one can also easily get the Casimir energy density by using the same approach used above, and the result is given by
 \begin{eqnarray}
 \nonumber
	&& E^{D}_{R}(a, m) =   -\frac{m^{D+1} \Gamma\left(-\frac{D+1}{2}\right)}{2^{D+2}\pi^{\frac{D+1}{2}}} \\
	& & \qquad - \sum_{n_{k}=-\infty}^{\infty}{}^{'}\left[\frac{ m }{2\pi f(n_{k},a_{k})}\right]^{\frac{D+1}{2}} \cos(2\pi  \mathbf{n}_{k}\cdot \mathbf{\theta}_{k} K_{\frac{D+1}{2}}\bigg( mf(n_{k},a_{k}) \bigg ) \,,
\end{eqnarray}
where we have defined the function
\begin{equation}
f(n_{k},a_{k}) = \sqrt{ \sum_{k=1}^{q} n_{k}^{2}a_{k}^{2}} \,.
\end{equation}

\section{Conclusion and discussion}
In conclusion, we have studied the Casimir effect of a scalar field under  the quasi-periodic boundary condition (\ref{bpcond}) with and without mass.  We find that  the Casimir force could be attractive or repulsive depending on values of the phase angle and especially, when the angle takes a  particular value, the Casimir effect disappears.  Further more, in arbitrary dimensions of spacetime,  the Casimir force gets a maximum value when the phase angle $\theta = 1/2$, which corresponds to anti-periodic boundary condition.

The exact value of $\theta$ may be dependent on the properties of the materials that are mimicked by the scalar field. So that the Casimir force could be attractive or repulsive depending on which kind of materials is used in the experiment. In another way, $\theta$ could be used to characterize whether the real material is perfect periodic or not. So, the effect from the phase angle  is worth further studying  and we suggest to do the experiment to verify our results.

\acknowledgments
This work is supported by National Science Foundation of China grant Nos.~11105091 and~11047138, ``Chen Guang" project supported by Shanghai Municipal Education Commission and Shanghai Education Development Foundation Grant No. 12CG51, National Education Foundation of China grant  No.~2009312711004, Shanghai Natural Science Foundation, China grant No.~10ZR1422000, Key Project of Chinese Ministry of Education grant, No.~211059,  and  Shanghai Special Education Foundation, No.~ssd10004, and the Program of Shanghai Normal University (DXL124).  

\appendix

\section{Regularization Type I }\label{reg1}
By using the Mellin transformation, we get
\begin{eqnarray}
\nonumber
	&& \pi^{-s/2}\Gamma\left(\frac{s}{2}\right) \,  \sum_{n=-\infty}^{\infty}(n+\nu)^{-s} \\
\nonumber
	&=& \sum_{n=-\infty}^{\infty} \int_{0}^{\infty} x^{\frac{s}{2}-1} e^{-(n+\nu)^{2}\pi x} dx \\
\nonumber
	&=& \int_{0}^{\infty} x^{\frac{s}{2}-1} e^{-\pi \nu^{2} x}\sum_{n=-\infty}^{\infty} e^{- \pi n^{2} x -2 \pi n \nu x} dx  \\
	&=& \int_{0}^{\infty} x^{\frac{s}{2}-1} e^{-\pi \nu^{2} x} \vartheta( i\nu x, ix ) dx \,,
\end{eqnarray}
where
\begin{equation}
	\vartheta(z, \tau) = \sum_{n=-\infty}^{\infty} e^{\pi i n^{2} \tau + 2\pi i n z}
\end{equation}
is the theta function, which satisfies
\begin{equation}\label{theta}
	\vartheta \left(z/\tau, -1/\tau \right) = (-i\tau)^{1/2} e^{i\pi z^{2}/ \tau} \vartheta(z, \tau)\,.
\end{equation}
Therefore, we have \footnote{Eq.(\ref{reflect}) will reduce to the usual reflection relation for the Riemann $\zeta(s)$ function by setting $\nu=0$.}
\begin{eqnarray} \label{reflect}
\nonumber
	&& \pi^{-s/2}\Gamma\left(\frac{s}{2}\right) \, \sum_{n=-\infty}^{\infty}(n+\nu)^{-s} \\
\nonumber
	&=& \int_{0}^{\infty} x^{\frac{s-3}{2}}  \vartheta( \nu , i/x ) dx
	  =  \int_{0}^{\infty}y^{\frac{1-s}{2} -1}  \vartheta( \nu , i y ) dy \\
\nonumber
	&=& \sum_{n=-\infty}^{\infty} e^{2\pi i n \nu} \int_{0}^{\infty}y^{\frac{1-s}{2} -1} e^{- \pi  n^{2} y  }  dy \\
	&=& \pi ^{-\frac{1}{2}+\frac{s}{2}} \Gamma\left(\frac{1-s}{2}\right)2\sum_{n=1}^{\infty} n^{s-1} \cos(2\pi  n \nu)
	\,,
\end{eqnarray}
or
\begin{eqnarray}
\nonumber
	&& \pi^{-s/2}\Gamma\left(\frac{s}{2}\right)\sum_{n=-\infty}^{\infty}(n+\nu)^{-s} \\
\nonumber
	&=& \pi ^{-\frac{1}{2}+\frac{s}{2}} \Gamma\left(\frac{1-s}{2}\right) \frac{ \partial^{s-1}\vartheta(\nu, 0) }{(2\pi i)^{s-1} \partial \nu ^{s-1}} \,,
\end{eqnarray}
for $s \geq 1$.

\section{Regularization Type II }\label{reg2}
By using the Mellin transformation again, we get
\begin{eqnarray}
\nonumber
	&& \pi^{-s/2}\Gamma\left(\frac{s}{2}\right) \,  \sum_{n=-\infty}^{\infty}  \bigg[ (n+\nu)^{2} + \mu^{2}\bigg]^{\frac{-s}{2}} \\
\nonumber
	&=& \sum_{n=-\infty}^{\infty} \int_{0}^{\infty} x^{\frac{s}{2}-1} e^{-[(n+\nu)^{2}+\mu^{2}]\pi x} dx \\
\nonumber
	&=& \int_{0}^{\infty} x^{\frac{s}{2}-1} e^{-\pi (\nu^{2} + \mu^{2}) x}\sum_{n=-\infty}^{\infty} e^{- \pi n^{2} x -2 \pi n \nu x} dx  \\
	&=& \int_{0}^{\infty} x^{\frac{s}{2}-1} e^{-\pi (\nu^{2} + \mu^{2}) x} \vartheta( i\nu x, ix ) dx \,.
\end{eqnarray}
After using eq.~(\ref{theta}), we have
\begin{eqnarray} \label{reflect2}
\nonumber
	&& \pi^{-s/2}\Gamma\left(\frac{s}{2}\right) \, \sum_{n=-\infty}^{\infty} \bigg[ (n+\nu)^{2} + \mu^{2}\bigg]^{\frac{-s}{2}} \\
\nonumber
	&=& \int_{0}^{\infty} x^{\frac{s-3}{2}}  e^{-\pi  \mu^{2} x} \vartheta( \nu , i/x ) dx \\
\nonumber
	&=&  \int_{0}^{\infty}y^{\frac{1-s}{2} -1}  e^{-\pi  \mu^{2}/y} \vartheta( \nu , i y ) dy \\
\nonumber
	&=& \sum_{n=-\infty}^{\infty} e^{2\pi i n \nu} \int_{0}^{\infty}y^{\frac{1-s}{2} -1} e^{- \pi  (n^{2} y + \mu^{2}/y)  }  dy \\
	&=& \mu^{1-s}\Gamma\left(\frac{s-1}{2}\right) + 4 \sum_{n=1}^{\infty}\left(\frac{\mu}{n}\right)^{\frac{1-s}{2}}  \cos(2\pi  n \nu) K_{\frac{1-s}{2}}(2\pi n\mu)	\,,
\end{eqnarray}
where $K_{\nu}(z)$ is the Bessel function. Here we have used
\begin{equation}
	\int_{0}^{\infty} x^{\nu-1} e^{-\gamma x - \frac{\beta}{x}} dx = 2 \left( \frac{\beta}{\gamma} \right)^{\nu/2}K_{\nu}(2\sqrt{\beta\gamma}) \,.
\end{equation}

\section{Odd values of  $D=2j+1$} \label{ovd}
By using the series representation of the Bernoulli polynomials
\begin{eqnarray}
	\varphi_{2n}(x) &=& \frac{(-1)^{n-1}2(2n)!}{(2\pi)^{2n}} \sum_{k=1}^{\infty} \frac{\cos(2k\pi x)}{k^{2n}} \,, \\
\nonumber
	&& (0\leq x \leq 1 \,, n = 1,2, \cdots) \,,
\end{eqnarray}
then, eq.~(\ref{vEnergy}) becomes
\begin{equation}
	E^{2j+1}_{R}(a) =  \frac{(-1)^{j+1}\pi^{j+1}}{2(j+1)a^{2(j+1)}} \frac{2^{2j+1}\Gamma(j+1)}{\Gamma(2j+2)} \varphi_{2(j+1)}(x)\,,
\end{equation}
where we have used $\Gamma(z+1)=z!$.  As we known, that
\begin{eqnarray}
	2^{2z-1}\Gamma(z)\Gamma\left(z+\frac{1}{2}\right) &=& \pi^{1/2} \Gamma(2z) \,, \\
	\Gamma(z)\Gamma(1-z) &=& \frac{\pi}{\sin(\pi z)} \,,
\end{eqnarray}
then, we get
\begin{equation}\label{eqgamma}
 	\frac{2^{2j+1}\Gamma(j+1)}{\Gamma(2j+2)} = (-1)^{j+1}\Gamma\left(-j-\frac{1}{2}\right) \pi^{-1/2}  \,.
\end{equation}
Inserting the above equation into eq.~(\ref{vEnergy}), we finally get the energy density with $D=2j+1$, i.e. eq.~(\ref{vEnergyodd}). Actually, an alternative way to get the expression (\ref{vEnergyodd}) is rather simple and straightforward.  From eq.~(\ref{energys}), we have
\begin{equation}
  E_{R}^{2j+1}(a) = -\frac{\pi^{j+\frac{1}{2}} \Gamma\left(-j -\frac{1}{2}\right)}{2a^{D+1}}\bigg[\zeta(-2j-1,\alpha)
  +\zeta( -2j-1, 1-\alpha)  \bigg]
\end{equation}
where $\zeta(s, \nu)$ is the Hurwitz $\zeta$ function.
By using  $\zeta(-n,x)=-\varphi_{n+1}(x)/(n+1)$, and $\varphi_{n}(1-a) = (-)^{n}\varphi_{n}(a)$,
one can also get eq.~(\ref{vEnergyodd}). So, our calculation is reliable.

\end{document}